\documentclass[reprint,twocolumn,aps,superscriptaddress]{revtex4-1}
\usepackage[utf8]{inputenc}
\usepackage[T1]{fontenc}
\usepackage{amsmath}
\usepackage{amsfonts}
\usepackage{physics}
\usepackage[caption=false]{subfig}
\usepackage{bm}
\usepackage{amssymb}
\usepackage[mathscr]{eucal}
\usepackage{graphicx}
\usepackage[dvipsnames]{xcolor}
\usepackage[hidelinks,colorlinks=true,allcolors=Blue]{hyperref}

\begin{document}
\title{Precision Franck-Condon spectroscopy from highly-excited vibrational states}
\author{Sindhana Pannir-Sivajothi}
\affiliation{Department of Chemistry and Biochemistry, University of California
San Diego, La Jolla, California 92093, USA}
\author{Joel Yuen-Zhou}
\email{joelyuen@ucsd.edu}

\affiliation{Department of Chemistry and Biochemistry, University of California
San Diego, La Jolla, California 92093, USA}

\begin{abstract}
As per the Franck-Condon principle, absorption spectroscopy reveals changes in nuclear geometry in molecules or solids upon electronic excitation. It is often assumed these changes cannot be resolved beyond the ground vibrational wavefunction width ($\sqrt{\hbar/m\omega}$). Here, we show this resolution dramatically improves with highly-excited vibrational initial states (with occupation number $\langle n\rangle$). These states magnify changes in geometry by $2\langle n\rangle +1$, a possibly counterintuitive result given the spatial uncertainty of Fock states grows with $n$. We also discuss generalizations of this result to multimode systems. Our result is relevant to optical spectroscopy, polariton condensates, and quantum simulators (\textit{e.g.}, boson samplers).
\end{abstract}
\maketitle
One of the standard paradigms for molecular and solid-state spectroscopy is the displaced harmonic oscillator (DHO) \cite{tokmakoff2014time,mukamel1995principles,fitchen1968physics,lax1952franck,huang1950theory}. This model contains a pair of electronic states -- the ground, $\ket{G}$, and the excited, $\ket{E}$, --  and one vibrational degree of freedom modeled as a harmonic oscillator. The equilibrium geometry of this oscillator determines the bond length of the molecule in a given electronic state and differs between the electronic states $\ket{G}$ and $\ket{E}$ by $\Delta d$. The Hamiltonian for this system is
\begin{equation}\label{eq:H}
	\begin{aligned}
		\hat{H}_0=&\ket{E}\bra{E}\hat{H}_{E}+\ket{G}\bra{G}\hat{H}_{G}\\
		=&\ket{E}\bra{E}\Big[\hbar\Omega+\hbar\omega_{a}(\hat{a}^{\dagger}-\sqrt{S})(\hat{a}-\sqrt{S})\Big]\\
		&+\ket{G}\bra{G}\hbar\omega_{a}\hat{a}^{\dagger}\hat{a},
	\end{aligned}
\end{equation}
where $\Omega$ is the frequency of the electronic transition, $\omega_{a}$ and $\hat{a}^{\dagger}$ are the frequency and creation operator of the vibrational mode. The Huang-Rhys factor $S$ and reorganization energy  $\hbar\omega_{a}S$ are the number of vibrational quanta and the corresponding energy absorbed in a vertical transition, respectively. Clearly, the reorganization energy can also be written as $\hbar\omega_{a}S=\frac{1}{2}m\omega_{a}^2(\Delta d)^2$, or in terms of the non-dimensional quantity $\Delta d^* = \sqrt{\frac{m\omega_{a}}{\hbar}}\Delta d$ as $S=\frac{1}{2}(\Delta d^*)^2$. Here, the vibrational eigenstates of $\hat{H}_G$ ($\hat{H}_E$) are labeled $\ket{n}$ ($\ket{n}'$).

\begin{figure*}[!t]
	\includegraphics[width=\linewidth]{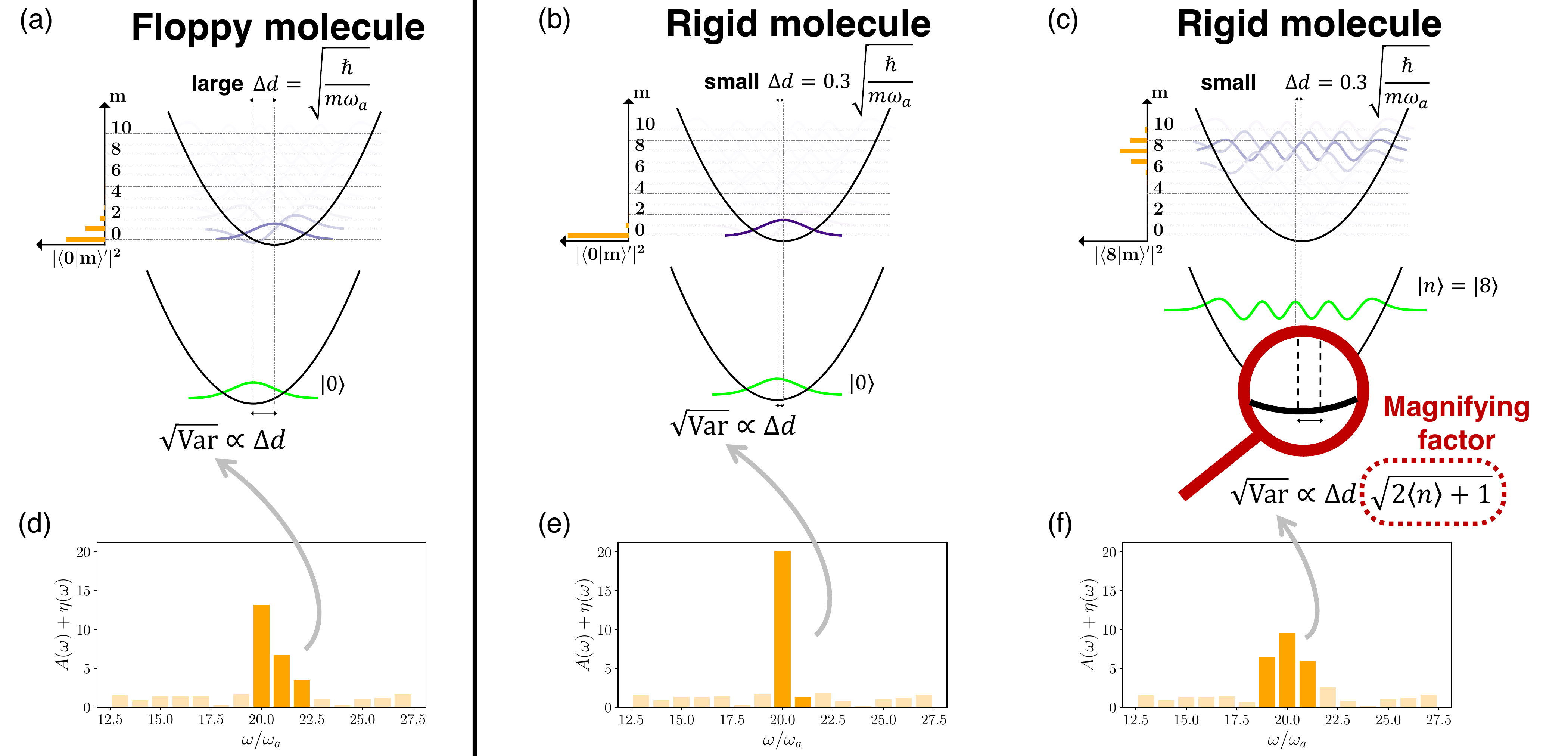}
	\caption{\label{fig:intuition} \textbf{Spectrum of a DHO with different initial states and displacements $\Delta d$.} (a-c)  Ground and excited potential energy surfaces are represented by two parabolas where changes in nuclear geometry, $\Delta d$, correspond to those of a floppy molecule $\Delta d = \sqrt{\frac{\hbar}{m\omega_a}}$ in (a) and a rigid one $\Delta d =0.3 \sqrt{\frac{\hbar}{m\omega_a}}$ in (b-c). The system is initially in the vibrational ground state $\ket{0}$ in (a-b) and in Fock state $\ket{8}$ in (c) with vibrational wavefunctions plotted in green. The orange bar plots depict Franck-Condon factors $|\bra{n}\ket{m}'|^2$; here, $\ket{m}'$ are the eigenstates of a harmonic oscillator displaced by $\Delta d$ from the ground state oscillator. Absorption cross-sections of (a-c) in the presence of noise $\eta(\omega)$ are shown in (d-f), with Var denoting the variance of $A(\omega)+\eta(\omega)$. Estimating $\Delta d$ from (b,e) becomes challenging due to the indistinguishability of its noisy spectrum from the $\Delta d=0$ spectrum. This can be improved by starting from a higher vibrational Fock state as shown in (c,f), which acts as a "magnifying lens" to determine the displacement $\Delta d$.}
\end{figure*}

 Typically in optical spectroscopy, the initial state of the system is assumed to be the ground state, $\ket{G}\ket{0}$ or a thermal state. The choice of ground state is a good approximation for high-frequency modes since, at room temperature, $k_BT\ll \hbar \omega_{a}$. Consequently, the Franck-Condon (FC) factors starting from $\ket{0}$,
\begin{equation}\label{eq:FC}
	|\bra{0}\ket{f}'|^2=e^{-S}\frac{S^f}{f!},
\end{equation}
are well understood textbook material \cite{tokmakoff2014time, mukamel1995principles,franck1926elementary,condon1928nuclear}. More generally, the evaluation of FC factors for multidimensional oscillators that feature Duschinky rotations and squeezing have also been extensively studied in the literature \cite{ansbacher1959note,chang2005new,ruhoff2000algorithms}. However, it is surprising that the use of higher vibrational Fock states as the initial state in optical spectroscopies has not been explored in detail. A few exceptions include precision measurements in diatomic molecules of electron-to-proton mass ratios \cite{zelevinsky2008precision,demille2008enhanced}, parity violation \cite{demille2008using}, and time-reversal symmetry violation \cite{kozyryev2017precision}. The present article aims to provide a comprehensive and intuitive analysis of their potential for highly-sensitive measurements of nuclear geometry. 

Hereafter, we theoretically study the electronic spectrum obtained when the initial vibrational occupation $\langle n \rangle > 0$ corresponds to a high-lying Fock, large-amplitude coherent, or a high-temperature thermal state, and find that it allows for the measurement of geometry changes at a resolution that is much improved compared to the vibrational vacuum. Importantly, we identify a length-scale associated with each Fock state $\ket{n}$ that determines such resolution. We also comment on a plausible generalization of these phenomena for multimode systems. Preparing these highly-excited initial vibrational states in a molecule can be an experimentally challenging task, although it is routinely done in gas phase spectroscopy of ultracold diatomic molecules through photoassociation schemes \cite{zelevinsky2008precision,demille2008enhanced}, and has been demonstrated \cite{weiner1990femtosecond} and proposed \cite{cina1990optical, banin1994impulsive, cina2000impulsive} in a few condensed-phase polyatomic systems using impulsive-stimulated Raman Scattering \cite{crimmins2002heterodyned}. 
In addition, a number of quantum simulation platforms that mimic the vibronic Hamiltonian of Eq. \ref{eq:H} such as boson samplers \cite{huh2015boson} and ion traps \cite{wolf2019motional,gilmore2021quantum, valahu2023direct} have emerged that can be initialized in well-defined vibrational states. Finally, molecular polariton condensates are also an ideal testbed for our findings, given the large occupations of the polariton modes associated with the onset of condensation \cite{pannir2022driving}.

Importantly, we note that the precision measurements we hereby discuss do not fall within the realm of \textit{quantum metrology}, which involves the use of quantum mechanical states to achieve higher levels of precision in measurements than possible with purely classical (coherent) states \cite{giovannetti2011advances}. An important goal of such field is to go beyond the standard quantum limit (SQL) (where measurement uncertainties scale as $1/\sqrt{N_{\mathrm{m}}}$ with number of measurements, $N_{\mathrm{m}}$), and reach the Heisenberg limit (where uncertainties scale as $1/N_{\mathrm{m}}$) \cite{giovannetti2011advances}. Previous studies have explored displacement measurements using Fock states that surpass the SQL \cite{wolf2019motional,gilmore2021quantum}. Notably, we stay within the SQL in our measurement scheme; as a result, large-amplitude coherent states, which are very classical states, can achieve similar levels of measurement precision as Fock states.

Information about nuclear geometry changes between electronic states is encoded in a molecule's spectrum. For a DHO, the absorption cross-section, 
\begin{equation}
	\begin{aligned}
		A(\omega)=&\Big(\frac{4\pi^2|\mu_{EG}|^2\omega}{\epsilon_0\hbar c}\Big)\\
		&\sum_{n=0}^{\infty}\rho_{n,n}\sum_{f=-n}^{\infty}|\bra{n}\ket{n+f}'|^2\delta(\omega-\Omega-f\omega_{a}),
	\end{aligned}
\end{equation}
when the initial vibrational density operator $\hat{\rho}=\sum_n\rho_{n,n}\ket{n}\bra{n}$ and the molecule's transition dipole moment is $\mu_{GE}$. This expression is derived by adding a light-matter coupling term $\hat{V}(t)=-\varepsilon(t)(\mu_{GE}\ket{G}\bra{E}+\mu_{EG}\ket{E}\bra{G})$ to the bare Hamiltonian $\hat{H}_0$ and applying first-order time-dependent perturbation theory \cite{tokmakoff2014time, tannor2007introduction}. Here, $\varepsilon(t)=\varepsilon_0\cos\omega t$ represents the electric field of a monochromatic electromagnetic wave probing the system. The Huang-Rhys parameter can be obtained from $A(\omega)$ by analyzing the moments of the probability density
\begin{equation}\label{eq:Pw}
	P(\omega)=\sum_{n=0}^{\infty}\rho_{n,n}\sum_{f=-n}^{\infty}|\bra{n}\ket{n+f}'|^2\delta(\omega-f\omega_{a}),
\end{equation}
termed the shifted FC weighted density of states \cite{huh2011application}.

Typically, $S$ is extracted from experimental data using the mean change in the number of vibrational quanta, obtained from the Stokes shift \cite{zuo2022dilution,de2015resolving}, or the full-width-half-maximum of the vibronic spectrum \cite{luo2018efficient,stadler1995optical,fitchen1968physics}. Both these approaches rely on expressions derived in 1952 by Lax  for the mean $\langle f \rangle$ and variance $	\langle(f-\langle f \rangle)^2 \rangle$ of the change in vibrational occupation \cite{lax1952franck}
\begin{subequations}\label{eq:Lax}   
\begin{align}
    \begin{split}\label{eq:Lax1}
        \langle f \rangle =& S
    \end{split}\\
    \begin{split}\label{eq:Lax2}
        \langle(f-\langle f \rangle)^2 \rangle=& S(2\langle n\rangle + 1).
    \end{split}
\end{align}
\end{subequations}
Note that the mean is independent of the initial vibrational state and the variance only depends on its occupation number $\langle n\rangle$. In typical room-temperature experiments, $\langle n \rangle \sim 0$ for high-frequency modes. Here, we will examine consequences of large $\langle n \rangle$.

A time-domain analysis offers a simple way to derive Eq. \ref{eq:Lax} and also provides crucial insights that help generalize our measurement scheme to multimode systems. The Fourier transform of the dipole correlation function $\langle \hat{\mu}(t)\hat{\mu}(0)\rangle =\langle e^{i\hat{H}_G t/\hbar}e^{-i\hat{H}_E t/\hbar} \rangle= e^{-i\Omega t}F(t)$ yields the absorption spectrum \cite{heller1981semiclassical,heller2018semiclassical}. Therefore, $F(t)$ acts as the characteristic function of $P(\omega)$ and the central moments in Eq. \ref{eq:Lax} can be obtained from $\langle f^k\rangle=\frac{1}{(\omega_a)^k}\frac{d^k}{d(-it)^k}F(t)|_{t=0}$ with $k=1,2$ \cite{lax1952franck}. Note that our definition of characteristic function $F(t)=\int_{-\infty}^{\infty}d\omega e^{-i\omega t}P(\omega)$ differs from the standard one by a minus sign in the exponent. For a DHO, the well-known expression
\begin{equation}\label{eq:Ft}
	F(t)=\exp\Bigg\{S\Big[(\langle n\rangle+1)(e^{-i\omega_a t}-1)+\langle n\rangle(e^{i\omega_at}-1)\Big]\Bigg\}.
\end{equation}
is derived using a second-order cumulant expansion and is exact for an initial thermal vibrational state $\hat{\rho}=e^{-\beta \hat{H}_G}/Z$ \cite{mukamel1995principles}; here, $\beta$ denotes the inverse temperature, and $Z$ is the partition function. On the other hand, when the system is initially in a Fock state $\ket{n}$, the first and second moments agree well with those calculated from Eq. \ref{eq:Ft} (equalling Eq. \ref{eq:Lax}), however, higher moments are significantly different.

In the absence of noise, it is possible to extract arbitrarily small $S$ with infinite precision from $A(\omega)$ using Eq. \ref{eq:Lax2}, regardless of the initial vibrational state. However, our objective is to quantify the precision limits when noise is present, so we add white noise $\eta(\omega)$, sampled from a uniform distribution over $[0,\eta_0]$ at each frequency $\omega$, to the absorption cross-section $A(\omega)$. When $S \gtrsim 1$, the Huang-Rhys parameter can be readily measured even from a noisy absorption spectrum, as depicted in Fig. \ref{fig:intuition}(a,d). However, when $S\ll 1$, due to the large overlap $|\bra{0}\ket{0}'|^2\approx 1$, the heights of the vibrational sidebands $|\bra{0}\ket{1}'|^2$ are negligible and might be indistinguishable from noise (see Fig. \ref{fig:intuition}(b,e) and Eq. \ref{eq:FC}). Consequently, the absorption spectrum for $S\ll 1$ closely resembles the $S=0$ case, making it challenging to accurately determine the value of $S$.

This problem can be overcome by initiating measurements from higher vibrational Fock states (see Fig. \ref{fig:intuition}c). For $n\gg 1$, the Franck-Condon factors for a DHO can be approximated as
\begin{equation}\label{eq:probapprox}
	|\bra{n}\ket{n+f}'|^2\approx e^{-(S+f)}(1+f)\frac{(Sn)^f}{(f!)^2}
\end{equation}
when $0\le f\ll n$ and $Sn< 1$ (Supplementary note 1). Comparison of this equation with the textbook Eq. \ref{eq:FC} reveals the change $S^{f} \to (Sn)^{f}$ when the initial vibrational state is a highly-excited one. Consequently, a small $S$ can be effectively enhanced to $Sn$; this magnifies the height of the vibrational sideband $|\bra{n}\ket{n+1}'|^2$, making it larger than noise even for $S\ll 1$ (see Fig. \ref{fig:intuition}(c,f)) and allows accurate extraction of small $S$ from a noisy spectrum.

\begin{figure}[!b]
	\includegraphics[width=\columnwidth]{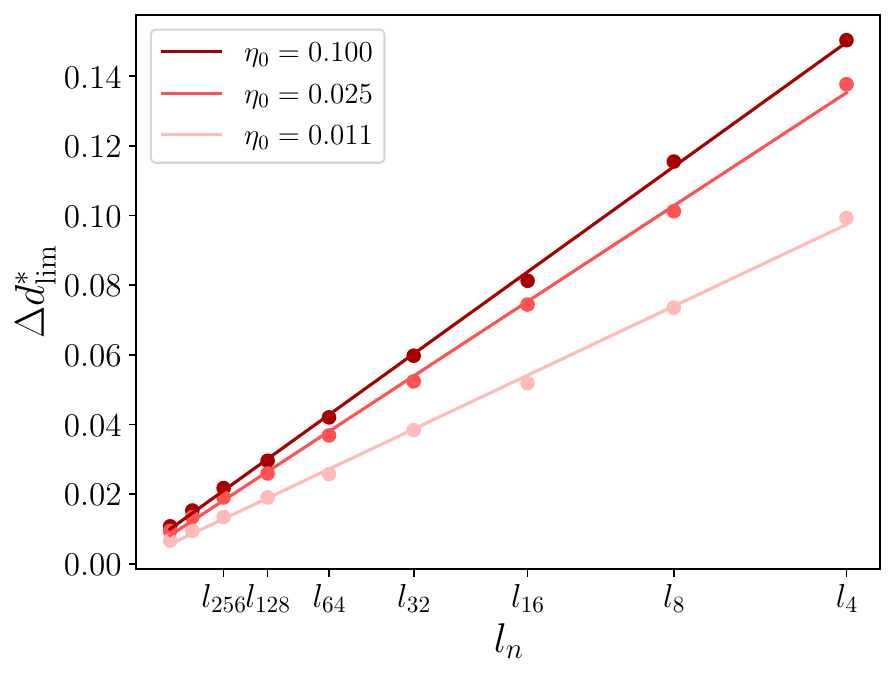}
	\caption{\label{fig:linear} \textbf{Precision limit for displacement measurements with Fock states.} The smallest displacement between the ground and excited potential energy surfaces, $\Delta d_{\mathrm{lim}}^*$, that can be measured with $<10\%$ error starting from different initial vibrational Fock states $\ket{n}$ is plotted against the `wavelength', $l_n\approx 4\sqrt{\frac{2\hbar}{m\omega_{a}}}\frac{1}{\sqrt{n}}$, associated with each Fock state for different values of the noise parameter $\eta_0$. We find a linear relationship between the precision limit and the wavelength of the vibrational Fock state used for the measurement. Here, $\eta_0$ is specified in units of $\Big(\frac{4\pi^2|\mu_{EG}|^2\Omega}{\epsilon_0\hbar c}\Big)$.}
\end{figure}

An alternate, physically intuitive interpretation of this magnifying effect involves the displacement $\Delta d$ rather than $S$. In Fig. \ref{fig:intuition}b, the change in geometry between electronic states, $\Delta d$, is notably smaller than the width, $ \sqrt{\frac{\hbar}{m\omega_a}}$, of the ground vibrational wavefunction $\psi_0(x)$. The natural length-scale $\sqrt{\frac{\hbar}{m\omega_a}}$ of the ground vibrational wavefunction "blurs" the real-space resolution for geometry change. This motivates us to define a length scale related to the initial vibrational state which dictates the smallest $\Delta d$ that can be measured using it. This length scale $l_{n}$, defined as $\frac{l_{n}}{2}:=\frac{2X_{n,\text{max}}}{n+1}$, corresponds to an average `wavelength' of the wavefunction $\psi_{n}(x)$ of Fock state $\ket{n}$. Here, $X_{n,\text{max}}$ refers to the classical turning point of a particle in a harmonic potential with a total energy of $E_{n}=\hbar\omega_a(n+1/2)$, satisfying $\frac{1}{2}m\omega_a^2X_{n,\text{max}}^2=E_{n}$. To define $l_n$, we divide $2X_{n,\text{max}}$ by the number of nodes, $n$, in $\psi_{n}(x)$, plus one. 

For large $n$, the length scale becomes $l_{n}\approx4\sqrt{\frac{2\hbar}{m\omega_a}}\sqrt{\frac{1}{n}}$. As $l_{n}$ decreases with increasing $n$, preparing the system in a Fock state with large $n$ may improve $\Delta d$ resolution;  the ratio $\Delta d/l_n$ becomes crucial here. In other words, this highly-excited vibrational state serves as a magnifying lens, enabling the detection of small changes in nuclear geometry. We numerically find a linear relationship between $l_n$ and the smallest displacement that can be measured (up to 10\% accuracy) $\Delta d_{\mathrm{lim}}$ with the corresponding Fock state (Fig. \ref{fig:linear}). This is analogous to how the smallest division in a ruler determines the smallest distance it can measure.

At first glance, our conclusion might seem counterintuitive given the large position uncertainty of $\psi_{n}(x)$, $\Delta x_{n} =\frac{1}{2}\sqrt{\frac{\hbar}{m\omega_a}}(2n+1)$. However, it is the length scale $l_{n}$ and not $\Delta x_{n}$ what serves as the ruler for nuclear geometry changes. Additionally, while our focus, for pedagogical reasons, has been on having an initial Fock state, Eq. \ref{eq:Lax2} states that the variance depends solely on the initial occupation number $\langle n \rangle$ of the mode and not on the specific initial state. Therefore, sensitivity enhancements by factors of $(2\langle n \rangle+1)$ on the measurement of the Huang-Rhys parameter should be possible with any state that has a large occupation number $\langle n \rangle$ and does not require a Fock state (Supplementary note 2).

\textit{Multimode.--} We generalize our measurement scheme to multimode systems by considering a system with $N+1$ vibrational modes coupled to an electronic transition. Our goal here is to measure the Huang-Rhys parameter $S_a$ of a specific mode with annihilation operator $\hat{a}$ and frequency $\omega_a$. The remaining modes have frequencies $\omega_{b_i}$, Huang-Rhys factors $S_{b_i}$, and annihilation operators $\hat{b}_i$ where $i\in\{1,..,N\}$. For simplicity, we restrict our analysis to displacements of modes, neglecting Duschinsky rotations and squeezing; this system is a spin-boson model. The Hamiltonian for this system is
\begin{equation}
	\begin{aligned}
		\hat{H}_0
		=&\ket{G}\bra{G}\Bigg[\hbar\omega_a\hat{a}^{\dagger}\hat{a}+\sum_{i=1}^N\hbar\omega_{b_i}\hat{b}_i^{\dagger}\hat{b}_i\Bigg]\\
		&+\ket{E}\bra{E}\Bigg[\hbar\Omega+\hbar\omega_a(\hat{a}^{\dagger}-\sqrt{S_a})(\hat{a}-\sqrt{S_a})\\
		&+\sum_{i=1}^{N}\hbar\omega_{b_i}(\hat{b}_i^{\dagger}-\sqrt{S_{b_i}})(\hat{b}_i-\sqrt{S_{b_i}})\Bigg].
	\end{aligned}
\end{equation}

The corresponding absorption cross-section
\begin{equation}	A(\omega)=\Big(\frac{4\pi^2|\mu_{EG}|^2\omega}{\epsilon_0\hbar c}\Big)\delta(\omega-\Omega)*\Big[P_{a}*P_{b_1}...*P_{b_{N}}\Big](\omega)
\end{equation}
involves the convolution of $N+1$ single mode probability densities $P_{a(b_i)}(\omega)$, given by Eq. \ref{eq:Pw} with the appropriate frequencies $\omega_a(\omega_{b_i})$, Huang-Rhys parameters $S_a(S_{b_i})$, and initial vibrational density operators $\hat{\rho}_{a}(\hat{\rho}_{b_{i}})$. As it is easier to factorize contributions of different modes from products rather than convolutions, we switch from frequency domain to time domain. The dipole correlation function $\langle \hat{\mu}(t)\hat{\mu}(0)\rangle = e^{-i\Omega t}F_{\mathrm{multi},\hat{\rho}_a}(t)$, where the multimode dephasing function $F_{\mathrm{multi},\hat{\rho}_a}(t)=F_{a,\hat{\rho}_a}(t)\prod_{i=1}^{N}F_{b_i,\hat{\rho}_{b_i}}(t)$ is a product of the single mode dephasing functions
\begin{equation}\label{eq:Fkrhok}
	\begin{aligned}
		F_{k,\hat{\rho}_{k}}(t)=&\sum_{n=0}^{\infty}\rho_{k,n,n}\sum_{f_{k}=-n}^{\infty} e^{-if_{k}\omega_{k}t}|\bra{n}_{k}\ket{n+f_{k}}_{k'}|^2,
	\end{aligned}
\end{equation}
with $k=a,b_i$. Here, the initial vibrational state is given by the factorizable density operator $\hat{\rho}_{a}\otimes_{i=1}^N\hat{\rho}_{b_{i}}$.

\begin{figure}[!b]
	\includegraphics[width=\columnwidth]{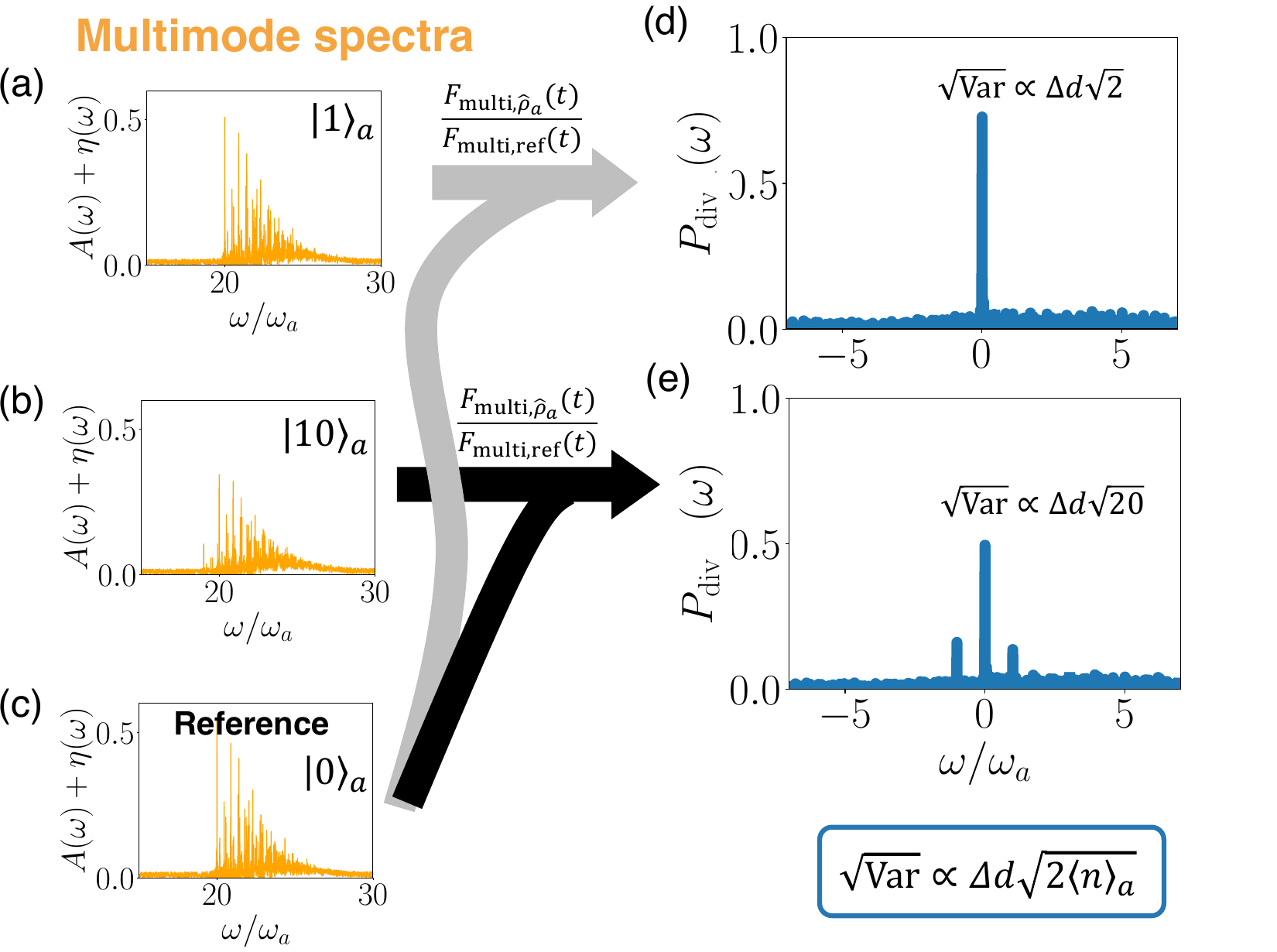}
	\caption{\label{fig:multimode}\textbf{Single-mode information from multi-mode spectra.} (a-c) The noisy absorption cross-section of a polyatomic molecule, $A(\omega)+\eta(\omega)$, with $\eta_0=0.001\Big(\frac{4\pi^2|\mu_{EG}|^2\Omega}{\epsilon_0\hbar c}\Big)$ starting from different initial vibrational Fock states, $\ket{n}_a$, in mode $a$ where $n=0,1,10$ are plotted in orange. The molecule has an electronic transition at frequency $\Omega=20\omega_{a}$ which is coupled to our vibrational mode of interest, mode $a$ (frequency, $\omega_{a}$, and displacement, $\Delta d_a^*$), and 10 other vibrational modes $b_i$ with frequencies, $\omega_{b_i}$, and displacements, $\Delta d_{b_i}^*$, chosen within $[0.2\omega_{a},2\omega_{a}]$ and $[0,\sqrt{2}]$, respectively (Supplementary note 4); for illustration, all modes $b_i$ are in the ground vibrational state initially $\prod_{i=1}^{10}\ket{0}_{b_i}$. (d-e) The Fourier transform of the divided time-domain signal which contains information only about mode $a$; this data-processing also works if the modes $b_i$ are not in their ground state.}
\end{figure}

Our aim is to isolate the contribution of mode $a$ to the spectrum. To do this, we keep the initial states $\hat{\rho}_{b_i}$ of all $b_i$ modes constant and vary only the state of $a$. We define a reference initial state $\hat{\rho}_{\mathrm{ref}}=\ket{0}_a\bra{0}_a\otimes_{i=1}^N\hat{\rho}_{b_{i}}$ which has dephasing function $F_{\mathrm{multi,ref}}(t)$, and use this to cancel out contributions of all modes other than $a$ as they are identical in $F_{\mathrm{multi,ref}}(t)$ and $F_{\mathrm{multi},\hat{\rho}_a}(t)$. We define the function $F_{\mathrm{div},\hat{\rho}_a}(t)=F_{\mathrm{multi},\hat{\rho}_a}(t)/F_{\mathrm{multi,ref}}(t)=F_{a,\hat{\rho}_a}(t)/F_{a,\mathrm{ref}}(t)$, where $F_{a,\mathrm{ref}}(t)$ is the $\hat{\rho}_a=\ket{0}_a\bra{0}_a$ case of the dephasing function $F_{a,\hat{\rho}_a}(t)$ in Eq. \ref{eq:Fkrhok}. Importantly, note that none of this depends on the vibrational state of the modes $b_i$. Keeping only up to the second order cumulant while calculating $F_{a,\hat{\rho}_a}(t)$ as in Eq. \ref{eq:Ft}, we get
\begin{equation}\label{eq:multi}
	\begin{aligned}
		F_{\mathrm{div},\hat{\rho}_a}(t)
		=&\exp\Big\{2S\langle n\rangle_a \Big[\cos(\omega_a t)-1\Big]\Big\}.
	\end{aligned}
\end{equation}

We use Cramer's criterion \cite{lukacs1960characteristic} to prove that $F_{\mathrm{div},\hat{\rho}_a}(t)$ is a characteristic function (Supplementary note 3) and define a probability density $P_{\mathrm{div},\hat{\rho}_a}(\omega)$ using its Fourier transform. Then, $P_{\mathrm{div},\hat{\rho}_a}(\omega)$ has mean and variance 
\begin{subequations}\label{eq:varSin}
	\begin{align}
            \begin{split}
                \langle  f_a\rangle=&0,
            \end{split}\\
		\begin{split}
		    \langle (f_a-\langle  f_a\rangle)^2 \rangle =&2S_a\langle n\rangle_a.
		\end{split}
	\end{align}
\end{subequations}
since $\langle f_a^k\rangle=\frac{1}{(\omega_a)^k}\frac{d^k}{d(-it)^k}F_{\mathrm{div},\hat{\rho}_a}(t)|_{t=0}$. As noted in the single mode case, the first and second moments agree well with Eq. \ref{eq:varSin} for all initial states $\hat{\rho}_a$, however, as Eq. \ref{eq:multi} is derived using a second-order cumulant expansion, higher moments differ for non-thermal states.
 
 In Fig. \ref{fig:multimode}d-e, we plot the probability density $P_{\mathrm{div},\hat{\rho}_a}(\omega)$ obtained in the presence of noise $\eta(\omega)$ in the reference spectrum (Fig. \ref{fig:multimode}c) and the vibrationally-excited spectra with $\hat{\rho}_a=\ket{1}_a\bra{1}_a$ (Fig. \ref{fig:multimode}a) and $\hat{\rho}_a=\ket{10}_a\bra{10}_a$ (Fig. \ref{fig:multimode}a) for a system that has $N=10$ modes (Supplementary note 4). Fig. \ref{fig:multimode}d-e look similar to the single mode probability densities extracted from Fig. \ref{fig:intuition}e-f. Furthermore, just like the single mode case, as the variance $2S_a\langle n\rangle_a$ of the probability density increases with the occupation $\langle n\rangle_a$ of the initial Fock state, their sensitivity towards measuring small $S_a$ also increases. Note how only a single peak is visible above the noise in Fig. \ref{fig:multimode}d and it will be indistinguishable from $P_{\mathrm{div},\hat{\rho}_a}(\omega)$ with $\hat{\rho}_a=\ket{0}_a\bra{0}_a$. Whereas, we see three peaks and a larger variance in Fig. \ref{fig:multimode}e. However, the tolerance to noise is much lower in the multimode case as noise will be present in both the numerator and denominator of $F_{\mathrm{div},\hat{\rho}_a}(t)$ and it becomes difficult to cancel out the contributions of the other $b_i$ modes.
 
Our initial recognition of the advantages associated with high-lying Fock states and large-amplitude coherent states came about while computing reaction rates afforded by polariton condensates which, by definition, involve macroscopic occupation of a polariton mode \cite{pannir2022driving}. In that context, the substantial number of excitations in the lower polariton, $N_{\mathrm{ex}}$, significantly enhances the polariton contributions to the rate constant -- from $1/N_{\mathrm{mol}}$ in the absence of condensation to $N_{\mathrm{ex}}/N_{\mathrm{mol}}$ in its presence, where $N_{\mathrm{mol}}$ is the number of molecules collectively coupled to a single photon mode. Similar results were obtained by Cortese \textit{et al} \cite{cortese2017collective}. The high Fock-state polariton Franck-Condon factors play a pivotal role in achieving this enhancement.

Diatomic molecules are the obvious candidates to test the presented ideas of measuring small changes in geometry upon electronic excitation, although the results above can only be trusted up to an upper bound in $n$, after which anharmonicities will alter our analysis \cite{zelevinsky2008precision,demille2008enhanced}. Interestingly enough, the vibrational modes of polyatomic molecules in condensed phase (\textit{e.g.}, in host matrices) can be very reasonably modelled as harmonic owing to central-limit theorem arguments \cite{makri1999linear,georgievskii1999linear,renger2002relation,pachon2014direct}, justifying the utility of spin-boson models in condensed phase chemistry \cite{nitzan2006chemical,mukamel1995principles,cao1996novel,reichman1996relaxation,egorov1998vibronic}. However, the challenge in this case is to prepare two initial states with drastically different occupations in a single mode of interest (see Eq. \ref{eq:multi}). It is conceivable how to achieve this if molecular parameters are well-known, but that defeats the purpose of the molecular geometry determination. Thus, we believe that in the short term, the most promising setups to test our ideas are quantum simulators such as trapped ions \cite{wolf2019motional,gilmore2021quantum, valahu2023direct} or quantum optics architectures optimized for boson sampling \cite{huh2015boson}. 

In summary, our present work has highlighted the unexplored potential of highly-occupied vibrational modes in magnifying the changes in equilibrium geometry during an electronic excitation, as detected in an optical absorption experiment. This potential can be tested in diatomic molecules in the harmonic regime or in quantum simulators for vibronic systems. Interestingly, it also plays an important role in the dynamics of polariton condensates. While our results are limited in practical scope given the harmonic assumption and the simplifications of the multimode analysis, we believe that they are very valuable as a reference to understand the essential physics of the problem. Future work involves generalizing the ideas of this article to include anharmonicities and mode-mixing effects.

S.P.-S. thanks Arghadip Koner and Yong Rui Poh for useful discussions. S.P.S. and J.Y.Z acknowledge funding support from the W. M. Keck Foundation. The code used in this work is available at  \textcolor{blue}{https://github.com/SindhanaPS/Franck-Condon\_spectroscopy\_measurement}.

\end{document}


\title{Supplementary information: Precision Franck-Condon spectroscopy from highly-excited vibrational states}
	
	\author{Sindhana Pannir-Sivajothi}
	\affiliation{Department of Chemistry and Biochemistry, University of California San Diego, La Jolla, California 92093, USA}
	\author{Joel Yuen-Zhou}
	\email{joelyuen@ucsd.edu}
	\affiliation{Department of Chemistry and Biochemistry, University of California San Diego, La Jolla, California 92093, USA}
	\maketitle
    \subsection*{Supplementary note 1: Franck-Condon factors involving Fock states $\ket{n}$ with $n\gg 1$}
    The Franck-Condon overlap is given by
    \begin{equation}\label{eq:FC}
    	\begin{aligned}
    		\bra{n}\ket{n+f}'=&(-1)^f\sqrt{\frac{(n+f)!}{n!}}\frac{\sqrt{e^{-S}S^f}}{f!}\sum_{m=0}^{n}\frac{(-S)^{m}f!}{(m+f)!}\binom{n}{m}
    	\end{aligned}
    \end{equation} 
    when $f\ge0$ \cite{palma1983franck}. We want to obtain an approximate expression for this overlap when $f\ll n$. To do this, we wish to truncate the summation over $m$. Let's define
    \begin{equation}
    	\begin{aligned}
    		c_{m}=&\frac{(-S)^{m}f!}{(f+m)!}\binom{n}{m}.
    	\end{aligned}
    \end{equation}
    Let's find the conditions under which the sequence $\{c_m\}$ decreases monotonically for all $f\ge 0$. If
    \begin{equation}
    	\begin{aligned}
    	\frac{|c_{m+1}|}{|c_m|}
    	=&\frac{S}{(f+m+1)}\frac{(n-m)}{(m+1)}<1
    	\end{aligned}
    \end{equation}
    for all $m$, then we have a decreasing sequence. This ratio $|c_{m+1}/c_m|$ is largest when $m=0$, therefore if
    \begin{equation}
    	\begin{aligned}
    		\frac{Sn}{(f+1)}<1
    	\end{aligned}
    \end{equation}
    it will definitely be a monotonically decreasing sequence. When $Sn<1$, the sequence will be monotonically decreasing for all $f\ge0$. Since it satisfies the alternating series test, the series $\sum_{m=0}^{n}c_m$ converges to $L$ which satisfies $|G_{k}-L|<|c_{k+1}|$ where $G_k=\sum_{i=0}^{k}c_{i}$. We focus on the case when $Sn < 1$; from the alternating series test, we see that the error is smaller than $|c_1|=Sn/(1+f)$ when we truncate the summation over $m$ at $m=0$. This gives us,
    \begin{equation}
    	\begin{aligned}
    		|\bra{n}\ket{n+f}'|^2=&\frac{(n+f)!}{n!}\frac{e^{-S}S^f}{(f!)^2}\Big(\sum_{m=0}^{n}\frac{(-S)^{m}f!}{(m+f)!}\binom{n}{m}\Big)^2\\
    		\approx&\sqrt{\frac{n+f}{n}}\frac{(n+f)^{n+f}}{n^n}e^{-f}e^{-S}\frac{S^f}{(f!)^2}\\
    		=&\Big(1+\frac{f}{n}
    		\Big)^{n+f+\frac{1}{2}}n^fe^{-f}e^{-S}\frac{S^f}{(f!)^2}\\
    		\approx&\Big[1+(n+f+\frac{1}{2})\frac{f}{n}
    		\Big]e^{-(S+f)}\frac{(Sn)^f}{(f!)^2}\\
    		\approx&e^{-(S+f)}(1+f)\frac{(Sn)^f}{(f!)^2}
    	\end{aligned}
    \end{equation} 
    where we have truncated the summation at $m=0$, used Stirling's approximation for $n!$ and $(n+f)!$ and taken $f\ll n$.

    \subsection*{Supplementary note 2: Comparison between Fock, coherent and thermal states}

\begin{figure}
	\includegraphics[width=0.5\columnwidth]{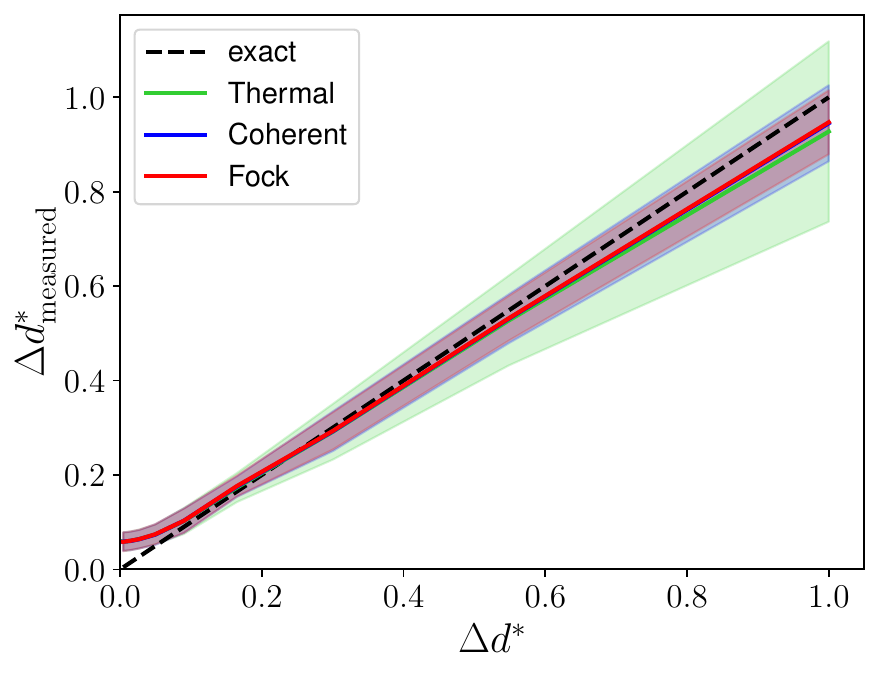}
	\caption{\label{fig:thermcoh} \textbf{Comparison between measured displacement for different initial states.} The displacement between the ground and excited potential energy surfaces, $\Delta d^*_{\mathrm{measured}}$, obtained from the variance of the spectrum in the presence of noise $\eta_0=0.1$ is plotted against the true value of $\Delta d^*$ that was used to generate the noise-free spectrum. This is done for different initial states with the same occupation number $\langle n\rangle =10$: thermal, coherent, and Fock. Here, $\eta_0$ is specified in units of $\Big(\frac{4\pi^2|\mu_{EG}|^2\Omega}{\epsilon_0\hbar c}\Big)$.}
\end{figure}

    Sensitivity enhancements by factors of $(2\langle n \rangle+1)$ on the measurement of the Huang-Rhys parameter should be possible with any state that has a large occupation number $\langle n \rangle$ and does not require a Fock state. To demonstrate this, we plot the mean $\Delta d_{\mathrm{measured}}$ averaged over many realizations of noise versus the true $\Delta d$ starting from different initial states with the same occupation number $\langle n \rangle=10$ in Fig. \ref{fig:thermcoh}. The thermal state, Fock state, and coherent state have similar mean $\Delta d_{\mathrm{measured}}$ but the standard deviation is larger for the thermal state. Fock states and coherent states perform similarly well, but thermal states perform slightly worse with larger standard deviations in the measured $\Delta d_{\mathrm{measured}}$ due to the longer tails of their probability density in frequency.

 \subsection*{Supplementary note 3: Characteristic function and Cramer's criterion}
   We use Cramer's criterion \cite{lukacs1960characteristic} which states
   \begin{quote}
       "A bounded and continuous function $f(t)$ is a characteristic function if, and only if, (i) $f(0)=1$ and (ii) $\psi(x,A)=\int_0^A\int_0^A f(t-u)\exp{ix(t-u)}dtdu$ is real and non-negative for all real $x$ and for all $A>0$."
   \end{quote}
    to prove that $F_{\mathrm{div},\hat{\rho}_a}(t)$ is a characteristic function. As
   $|F_{\mathrm{div},\hat{\rho}_a}(t)|\le 1$, it is bounded. $F_{\mathrm{div},\hat{\rho}_a}(t)$ is also continuous and satisfies $F_{\mathrm{div},\hat{\rho}_a}(0)=1$. To show that $F_{\mathrm{div},\hat{\rho}_a}(t)$ satisfies condition (ii) we expand
   \begin{equation}
   	\begin{aligned}
   		F_{\mathrm{div},\hat{\rho}_a}(t)
   		=&e^{-2Sn}e^{Sne^{i\omega_{a} t}}e^{Sne^{-i\omega_{a}t}}\\
   		=&e^{-2Sn}\Big[1+Sne^{i\omega_{a} t} +\frac{(Sn)^2}{2!}e^{i2\omega_{a} t}+...\Big]\Big[1+Sne^{-i\omega_{a} t} +\frac{(Sn)^2}{2!}e^{-i2\omega_{a} t}+...\Big]\\
     =&\sum_{m=-\infty}^{\infty}c_me^{im\omega_{a}t}.
   	\end{aligned}
   \end{equation}
   We note that $c_m\ge0$ for all $ m\in \mathbb{Z}$ and $\sum_{m=-\infty}^{\infty}c_m=1$ as $F_{\mathrm{div},\hat{\rho}_a}(0)=1$. The function 
   \begin{equation}
       \begin{aligned}
       \psi(x,A)=&\int_0^A\int_0^AF_{\mathrm{div},\hat{\rho}_a}(t-u) \exp[ix(t-u)]dtdu\\
       =&\int_0^A\int_0^A\Big(\sum_{m=-\infty}^{\infty}c_me^{im\omega_{a}(t-u)}\Big) \exp[ix(t-u)]dtdu\\
       =&\sum_{m=-\infty}^{\infty}c_m\int_0^Adte^{i(x+m\omega_{a})t}\int_0^Adue^{-i(x+m\omega_{a})u}\\
       =&\sum_{m=-\infty}^{\infty}c_m\Bigg[\frac{e^{i(x+m\omega_{a})A}-1}{i(x+m\omega_{a})}\Bigg]\Bigg[\frac{e^{-i(x+m\omega_{a})A}-1}{-i(x+m\omega_a)}\Bigg]\\
       =&\sum_{m=-\infty}^{\infty}c_m\Bigg[\frac{2-2\cos[(x+m\omega_{a})A]}{(x+m\omega_{a})^2}\Bigg]\\
       =&\sum_{m=-\infty}^{\infty}c_m\Bigg[\frac{4\sin^2[(x+m\omega_{a})A/2]}{(x+m\omega_{a})^2}\Bigg]
       \end{aligned}
   \end{equation}
   is non-negative for all real $x$ and $A>0$ as $c_m\ge 0$. We need to prove that $\psi(x,A)$ is finite. As $\Big[\frac{4\sin^2((x+m\omega_{a})A/2)}{(x+m\omega_{a})^2}\Big]\le \frac{4}{x^2}$, we can find a finite upper bound for $\psi(x,A)=\sum_{m=-\infty}^{\infty}c_m\Big[\frac{4\sin^2((x+m\omega_{a})A/2)}{(x+m\omega_{a})^2}\Big]\le \frac{4}{x^2}\sum_{m=-\infty}^{\infty}c_m=\frac{4}{x^2}$ for $x\neq 0$. When $x=0$, $\psi(x,A)=\sum_{m=-\infty}^{\infty}c_mA^2\Big[\frac{\sin^2(m\omega_{a}A/2)}{(m\omega_{a}A/2)^2}\Big]=\sum_{m=-\infty}^{\infty}c_mA^2\mathrm{sinc} ^2(m\omega_{a}A/2)\le A^2\sum_{m=-\infty}^{\infty}c_m=A^2$ is bounded. Hence, the function $F_{\mathrm{div},\hat{\rho}_a}(t)$ satisfies Cramer's criterion and is a characteristic function.

	\subsection*{Supplementary note 4: Multimode systems}
	The probability density $P(\omega)$ and dephasing function $F(t)$ are related through
	\begin{equation}
		F(t)=\int_{-\infty}^{\infty}d\omega e^{-i\omega t}P(\omega).
	\end{equation}
    The inverse transformation for a single-mode system gives  
    \begin{equation}
    	\begin{aligned}
     	    P(\omega)=&\frac{1}{2\pi}\int_{-\infty}^{\infty}dt e^{i\omega t}F(t)\\
    		=&\frac{1}{2\pi}\int_{-\infty}^{\infty}dt e^{i\omega t}\Big[\sum_{n=0}^{\infty}\rho_{n,n}\sum_{f=-n}^{\infty}|\bra{n}\ket{n+f}'|^2e^{-if\omega_a t}\Big]\\
    		=&\sum_{n=0}^{\infty}\rho_{n,n}\sum_{f=-n}^{\infty}|\bra{n}\ket{n+f}'|^2\delta(\omega-f\omega_{a})
    	\end{aligned}
    \end{equation}
    as $\delta(\omega)=\frac{1}{2\pi}\int_{-\infty}^{\infty}e^{i\omega t}dt$. 
    
    For a multimode system, convolution of $P_a(\omega)*P_{b_1}(\omega)... *P_{b_{N_{\mathrm{modes}}}}(\omega)$ leads to products in time domain. We will demonstrate the two mode case
    \begin{equation}
    	\begin{aligned}
    		P_a(\omega)*P_{b_1}(\omega)=&\int_{-\infty}^{\infty}d\omega'P_a(\omega')P_{b_1}(\omega-\omega')\\
    		=&\int_{-\infty}^{\infty}d\omega'\Bigg[\frac{1}{2\pi}\int_{-\infty}^{\infty}dt e^{i\omega' t}F_a(t)\Bigg]\Bigg[\frac{1}{2\pi}\int_{-\infty}^{\infty}dt' e^{i(\omega-\omega') t'}F_{b_1}(t')\Bigg]\\
    		=&\int_{-\infty}^{\infty}dt'\frac{1}{2\pi}\int_{-\infty}^{\infty}dt \frac{1}{2\pi}\int_{-\infty}^{\infty}d\omega'e^{i\omega' (t-t')}F_a(t) F_{b_1}(t')e^{i\omega t'}\\
    		=&\frac{1}{2\pi}\int_{-\infty}^{\infty}dtF_a(t)\int_{-\infty}^{\infty}dt'  \delta(t'-t)F_{b_1}(t')e^{i\omega t'}\\
    		=&\frac{1}{2\pi}\int_{-\infty}^{\infty}dte^{i\omega t}F_a(t)F_{b_1}(t),
    	\end{aligned}
    \end{equation}
	and we have
	\begin{equation}
		\begin{aligned}
			P_a(\omega)*P_{b_1}(\omega)
			=&\Bigg[\sum_{f_a=-n}^{\infty}|\bra{n}_a\ket{n+f_a}_{a'}|^2\delta(\omega-f_a\omega_a)\Bigg]*\Bigg[\sum_{f_{b_1}=0}^{\infty}|\bra{0}_{b_1}\ket{f_{b_1}}_{b_1'}|^2\delta(\omega-f_{b_1}\omega_{b_1})\Bigg]\\
			=&\sum_{f_a=-n}^{\infty}\sum_{f_{b_1}=0}^{\infty}|\bra{n}_a\ket{n+f_a}_{a'}|^2|\bra{0}_{b_1}\ket{f_{b_1}}_{b_1'}|^2\Big[\delta(\omega-f_a\omega_a)*\delta(\omega-f_{b_1}\omega_{b_1})\Big]\\
			=&\sum_{f_a=-n}^{\infty}\sum_{f_{b_1}=0}^{\infty}|\bra{n}_a\ket{n+f_a}_{a'}|^2|\bra{0}_{b_1}\ket{f_{b_1}}_{b_1'}|^2\delta(\omega-f_a\omega_a-f_{b_1}\omega_{b_1}),
		\end{aligned}
	\end{equation}
    where the convolution of a pair of delta functions is
    \begin{equation}
    	\begin{aligned}
    		\delta(\omega-f_a\omega_a)*\delta(\omega-f_{b_1}\omega_{b_1})=&\int_{-\infty}^{\infty}d\omega'\delta(\omega'-f_a\omega_a)\delta(\omega-\omega'-f_{b_1}\omega_{b_1})\\
    		=&\delta(\omega-f_a\omega_a-f_{b_1}\omega_{b_1}).
    	\end{aligned}
    \end{equation}
   Frequencies and Huang-Rhys parameters of the $N=10$ $b_i$ modes used in the calculation in Fig. 3 are
   \begin{equation}
   	\begin{aligned}
   		\omega_{b_i}/\omega_{a}\in&\{0.18, 0.46, 0.57, 0.86, 0.89, 1.36, 1.41, 1.52, 1.88, 2.06\},\\
   		S_{b_i}\in&\{0.25, 0.54, 0.41, 0.16, 0.75, 0.19, 0.65, 0.04, 0.11, 0.27\}.
   	\end{aligned}
   \end{equation}